# Jerky elasticity: Avalanches and the martensitic transition in $Cu_{74.08}\,Al_{23.13}\,Be_{2.79}$ shape-memory alloy

Ekhard K.H. Salje

*Department of Earth Sciences, University of Cambridge, Downing Street, Cambridge CB2 3EQ, UK*

Johannes Koppensteiner, Marius Reinecker, Wilfried Schranz

*Faculty of Physics, University of Vienna, Boltzmanngasse 5, A-1090 Vienna, Austria*

Antoni Planes

*Departament d'Estructura i Constituents de la Matèria. Facultat de Física. Universitat de Barcelona. Diagonal 647, E-08028 Barcelona, Catalonia*

Jerky elasticity was observed by Dynamical Mechanical Analyzer (DMA) measurements in a single crystal of the shape memory alloy $Cu_{74.08}\,Al_{23.13}\,Be_{2.79}$. Jerks appear as spikes in the dissipation of the elastic response function and relate to the formation of avalanches during the transformation between the austenite and the martensite phase. The statistics of the avalanches follows the predictions of avalanche criticality $P(E) \sim E^{-\varepsilon}$ where $P(E)$ is the probability of finding an avalanche with the energy E. This result reproduces, within experimental uncertainties, previous findings by acoustic emission techniques.

The elastic response of materials measured at low frequencies, such as measured in a Dynamical Mechanical Analyzer (DMA) arrangement, is generally found to be a smooth function of the applied stress in the regime of small stresses. Superelastic softening occurs when the external stress is involved in the movement of interfaces such as twin boundaries or the interface between two phases (such as a para-elastic and a ferroelastic phase [1]). The real part of the elastic response relates to the elastic moduli while the imaginary part measures dissipation. Previous studies of the dissipated energy revealed Debye type relaxations related to the movement of such interfaces [2-9]. Jerky elasticity, where the dissipation displays singularities such as spikes, was envisaged theoretically but has never been seen experimentally in DMA experiments. On the other hand, jerky behavior was observed in acoustic emission measurements [10-11] and it appears likely that equivalent measurements should be possible when the driving force of the interfacial movement is stress rather than strain or thermal. We will show below that we see, indeed, the same jerky behaviour as in acoustic emission measurements when appropriate DMA measurements are performed.

In general, for avalanches to occur it is required that thermal fluctuations play a minor role (athermal behavior) and that the system is driven slowly enough across the transition in order to ensure that avalanches do not overlap. On the other hand, a quantitative study of avalanches requires the use of very sensitive techniques that enable detection of very small transformed fractions characterized by a sub micrometer length scale and good time resolution. In the case of structural transitions, including ferroics and multiferroics, acoustic emission (AE) is a very convenient technique for the study of avalanches. In some cases, other more integral methods including calorimetry have been used to reveal the existence of avalanches [15]. It is worth pointing out that no avalanches have been detected so far using Dynamic Mechanical Analysis (DMA) measurements.

We have performed a DMA study of avalanches during the thermally induced martensitic transition in a $Cu_{74.08} Al_{23.13} Be_{2.79}$ single crystal [15]. This is a technologically interesting material which displays shape-memory and superelastic properties. The crystal has been grown by Bridgmann technique. The atomic composition of the sample is $Cu_{74.08} Al_{23.13} Be_{2.79}$ as determined by microprobe analysis of Cu and Al while the Be content was inferred from the composition of the melt [15]. In the absence of applied external forces, this system undergoes a martensitic transition on cooling to a monoclinic 18R structure in the range from (approx.) 260 K to 210 K. From the original rod, a sample with length 7.2 mm, width 1.05 mm and thickness 0.13mm has been cut using a low-speed diamond saw. The surface of the sample is parallel to (100) plane and its length parallel to [110] direction.

The DMA experiment was performed on a Perkin Elmer DMA Diamond instrument with small applied forces (<50mN applied force with a dc component of 60mN). The displacement of the sample was ca. 15μm. Increasing the forces did not lead to an increase of the spikes but simply deformed the sample macroscopically. The deformation mode was the three point bending method [3] under very slow heating and cooling conditions (< 0.15 K/hour). Increased heating and cooling rates lead to the overlap between the spikes so that any statistical evaluation of the spectrum of spikes became impossible. The applied frequency for the bending stress was also as low as was technically possible (0.1Hz). The temperature dependence of the dissipation (tan δ) during a heating experiment is shown in Fig.1, dynamic phase lag tan δ in Fig. 2. At the onset of the transition near 270K a smooth transition zone is visible while jerky elasticity is indicated by spikes at temperatures below 265K down to the lowest temperature measured in this experiment (205K). This interval is slightly larger than that measured in another sample cut from the same growth body and described in [15]. The profiles of the spikes are essential identical so that a statistical analysis is possible by simply taking the amplitudes of the spikes into consideration. The analysis of the results in Fig.2 is represented in Fig.3. An estimation of a probability density so that P(E)dE is the probability of finding an event within the range E to E+dE is given by P(E). The distribution P(E) in Fig.3 shows a maximum near zero amplitude with a total count of 44 events. The distribution decays to half this values at an amplitude of 0.55 and a small number of events at very large amplitudes. The functional form of the decay follows a power law (Fig.4) with an approximate exponent of ε= -1.3.

We can now compare this result with previous findings. Avalanches in martensitic transitions have commonly been studied from measurements of the AE generated during the transition. In this kind of measurements avalanches are detected at time scales (in the range from μs to ms) much smaller than in our present experiments and the number of detected avalanches is typically orders of magnitude larger. It must be pointed out that in AE measurements the avalanche size is often characterized from the amplitude A of the detected signals. Power law behavior is also obtained characterized by an exponent α (P(A) ~ $A^\alpha$). The energy exponent can then be obtained by assuming a statistical dependence E ~ $A^2$ between energy and amplitude which lead to ε = - (α +1)/2 [16]. For Cu-based shape memory alloys (including Cu-Al-Be) transforming to a 18R martensite, an exponent $\alpha = -3 \pm 0.2$ has been reported [17]. The corresponding energy exponent is $\varepsilon = -2 \pm 0.2$. This exponent is somewhat larger than our presenent DMA estimation . The analytical error in the determination of the exponent from our experiment can be estimated to be around 0.3 which leads to an upper bound of 1.6. While this value is still smaller than in [17] our power law interval for the energy bursts is only one decade so that further numerical systematic errors are possible. Nevertheless, this papers shows that the power law exponent can be measured directly by the surprisingly simple DMA technique and opens the way for further experimental studies of the power law exponent.

This work was supported by CICyT (Spain) Project No. MAT-2007-61200, Marie-Curie RTN MULTIMAT (Contract No. MRTN-CT-2004-505226)

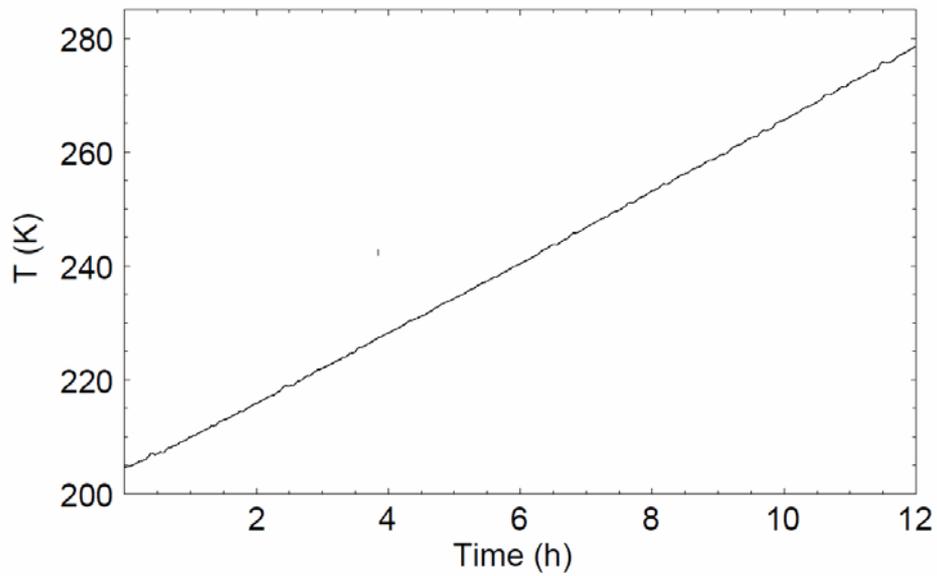

Fig. 1. T-ramp during measurement of Fig. 2 showing a linear ramp of 0.15 K/min driven for 12 h.

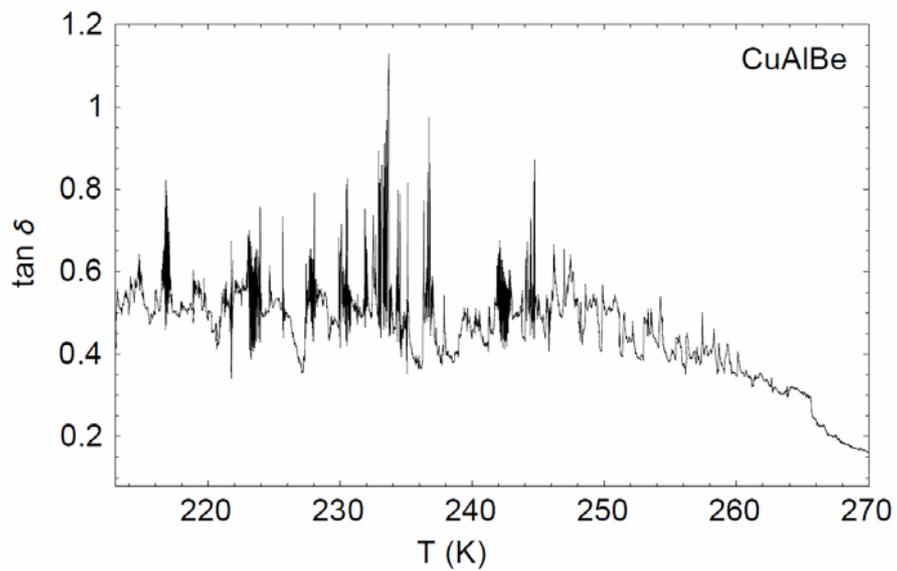

Fig. 2. Phase lag tan($\delta$) of a $Cu_{74.08} Al_{23.13} Be_{2.79}$ single crystal recorded in three-point-bending mode at 0.1 Hz and heating rate of 0.15 K/min.

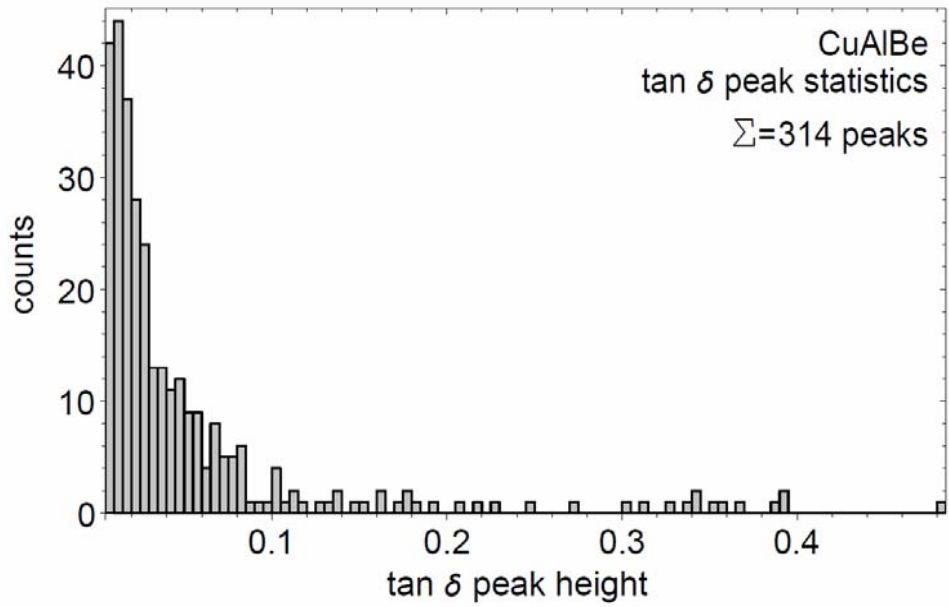

Fig. 3. Peak statistics of tan δ-peaks of Fig 1 classifying 314 peaks according to individual absolut height.

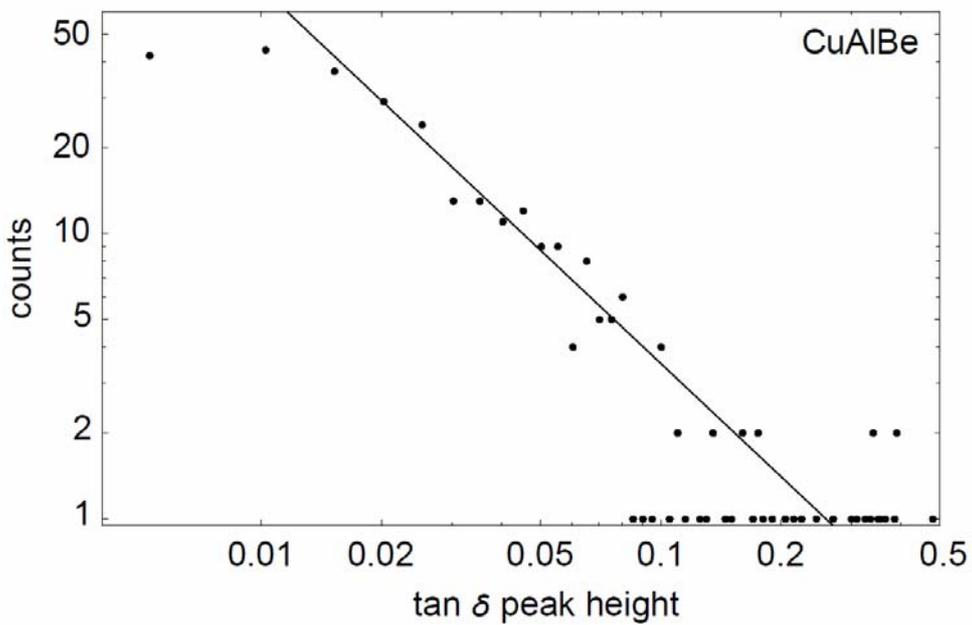

Fig. 4. Log-Log plot of peak statistics of Fig 3. Fitted line displays a power law dependence with an energy exponent of $\varepsilon = -1.3$ ($P(E) \sim E^{-1.3}$).

References

[1] E.K.H Salje, Phase transitions in ferroelastic and co-elastic crystals, Cambridge University Press. Cambridge UK (1993) see also E.K.H. Salje, Acta Crystallogr. A 47, 453 (1991).

[2] A.V. Kityk, W. Schranz, P. Sondergeld P, D. Havlik, E.K.H. Salje, J.F. Scott, Phys. Rev. B 61, 946 (2000).

[3] R.J. Harrison, S.A.T. Redfern, A. Buckley, E.K.H. Salje, J.Appl. Phys. 95, 1706 (2004).

[4] R.J. Harrison, S.A.T. Redfern, E.K.H. Salje, Phys. Rev, B 69, 144101 (2004).

[5] E.K.H. Salje, H.L. Zhang, J.Phys. Condens. Matter 21, 035901 (2009).

[6] E.K.H. Salje, Phys. Chem. Minerals 35, 321 (2008).

[7] E. SALJE, Phys. Chem. Minerals 15, 336 (1988).

[8] L. Goncalves-Ferreira, S.A.T. Redfern, E. Atacho E, E.K.H. Salje, Appl. Physics Lett. 94, 081903 (2009) see also W.Schranz,P.Sondergeld, A.V. Kityk, E.K.H.Salje, Phys. Rev. B 80, 094110 (2009).

[9] K. Trachenko, V.V. Brazhkin, O.B. Tsiok, M.T. Dove, E.K.H. Salje, Phys. Rev. B 76, 012103 (2007).

[10] E. Vives, J. Ortin, L.Manosa, I. Rafols, R. Perezmagrane, A. Planes, Phys. Rev. Lett. 72, 1694 (1994).

[11] L Carrillo, L.Manosa, J.Ortin, A. Planes, E. Vives, Phys. Rev. Lett. 81, 1889 (1998).

[12] F.J. Perez-Reche, M. Stipcich, E. Vives, L. Manosa, A. Planes, M. Morin, Phys. Rev. B 69, 064101 (2004).

[13] S. Sreekala, R. Ahluwalia, G. Ananthakrishna, Phys. Rev. B 70, 224105 (2004).

[14] F.J. Perez-Reche, L. Truskinovsky, G. Zanzotto, Continuum Mechanics &

Themodynamics 21, 17 (2009).

[15] E. K. H. Salje, H. Zhang, H. Idrissi, D. Schryvers, M. A. Carpenter, X. Moya, A. Planes, Phys. Rev. B 80 in press (2009).

[16]E. Bonnot, L. Mañosa, A. Planes, D. Soto-Parra, E. Vives, B. Ludwig, C. Strothkaemper, T. Fukuda, T. Kakeshita, Phys. Rev. B, **78**, 184103 (2003).

[17] L. Carrillo, L. Mañosa, J. Ortín, A. Planes, E. Vives, Phys. Rev. Lett., **81**, 1889 (1998).